\documentstyle[sprocl]{article}

\def\Journal#1#2#3#4{{#1} {\bf #2}, #3 (#4)}

\def\NCA{\em Nuovo Cimento}

\def\be{\begin{equation}}
\def\ee{\end{equation}}
\def\bea{\begin{eqnarray}}
\def\eea{\end{eqnarray}}
\def\bra#1{\langle #1 |}
\def\ket#1{| #1\rangle }

\begin{document}

{\hfill UGVA-DPT-09-1014}
\bigskip

\title{A THEORY OF ALGEBRAIC INTEGRATION}

\author{R. CASALBUONI\footnote{On leave from the Dept. of Physics,
University of Florence, Florence, Italy}}

\address{D\'epartement de Physique, Universit\'e de Gen\`eve,
Gen\`eve, Suisse\\E-mail: casalbuoni@fi.infn.it}


\maketitle\abstracts{
In this paper we study the problem of quantizing theories defined
over a nonclassical configuration space. If one follows the
path-integral approach, the first problem one is faced with is
the one of definition of the integral over such spaces. We
consider this problem and we show how to define an integration
which  respects the physical principle of composition of the
probability amplitudes for a very large class of algebras.}

\section{Introduction and motivations}

In the ordinary description of quantum mechanics the classical
phase space gets promoted to a noncommutative algebra. However,
in the path-integral quantization one still retains a classical
phase space (but with a different physical interpretation). The
situation changes in supersymmetric theories where one deals with
a phase space including Grassmann variables \cite{casalbuoni1}.
Correspondingly one faces the problem of defining the integration
over a nonclassical space. This of course is solved by means of
the integration rules for Grassmann algebras introduced by
Berezin \cite{berezin}. These rules were originally introduced in
relation to the quantization of  Fermi fields and they were
justified by the fact that they reproduce the perturbative
expansion. In a more general setting one would like to know if
there exists a principle underlying these rules. A possible
answer is the translational invariance of the Berezin's rules. In
fact the induced functional measure is again translational
invariant and this has as a consequence the validity of the
Schwinger quantum action principle. However, if one looks for
generalizations of ordinary quantum mechanics, as for instance
quantum groups, $M$-theory, etc, it is not clear whether one
would insist on the validity of such a principle. We would feel
more confident if we could rely upon something more fundamental.
Our proposal is to use as a fundamental property the {\em
combination law for probability amplitudes}. In the usual path
integral approach this is a trivial consequence of the
factorization properties of the functional measure, which in turn
is equivalent to the completeness of the position eigenstates. In
the general situation we are considering, the space of the
eigenvalues of the position operator goes into a non classical
space. By following the spirit of noncommutative geometry
\cite{connes} we will deal with this situation by considering the
space of the functions on the space, rather than the space itself
(which in general has no particular meaning). Therefore, one has
to know how to lift up the completeness relation from the space
to the space of the functions. We can see how things work in the
classical case. Consider an orthonormal set of states
$|\psi_n\rangle$, then we can convert the completeness relation
in the configuration space into the orthogonality relation for
the functions $\psi_n(x)=\langle x|\psi\rangle$
\be
\int\langle\psi_m|x\rangle\langle x|\psi_n\rangle=\int \psi_m^*(x)\psi_n(x)
dx=\delta_{nm}
\ee
Notice that the completeness in the $x$-space can be reconstructed from this
orthonormality relation and from the completeness of the states. The
properties of the set $\{\psi_n(x)\}$ following from the completeness are
\begin{itemize}
\item The set $\{\psi_n(x)\}$ spans a vector space.
\item The product $\psi_m(x)\psi_n(x)$ can be expressed as a linear
combination of elements of the set $\{\psi_n(x)\}$.
\end{itemize}
In other words, the set $\{\psi_n(x)\}$ has the structure of an algebra. By
using again the completeness we can write
\be
\psi_n^*(x)=\sum_m C_{nm}\psi_m(x)
\ee
from which
\be
\int \sum_m C_{nm}\psi_m(x)\psi_p(x) dx=\delta_{np}
\ee
In the following we will use this formula to define the integration over
an algebra with basis elements $\{x_i\}$
\be
\int_{(x)}\sum_jC_{ij}x_jx_k=\delta_{ik}
\ee
with $C$ a matrix to be specified.

\section{Algebras}

An algebra is a vector space $\cal A$ equipped with a bilinear
mapping ${\cal A}\times{\cal A}\to{\cal A}$. We will work in a
fixed basis $\{x_i\}$, with $i=0,\cdots,n$ (we will consider also
the possibility of $n\to\infty$, or of a continuous index). The
algebra product is determined by the structure constants
$f_{ijk}$
\be
x_ix_j=f_{ijk}x_k
\ee
An important tool is the algebra of the left and right
multiplications. The associated matrices are defined by
\be
R_a\ket x=\ket x a,~~~~~\bra x L_a=a\bra x,~~~~~a\in{\cal A}
\ee
with $\bra x=(x_0,x_1,\cdots,x_n)$. It follows ($L_i\equiv
L_{x_i}$)
\be
(R_i)_{jk}=f_{jik},~~~~~(L_i)_{jk}=f_{ikj}
\ee
These matrices encode all the properties of the algebra. For
instance, in the case of an associative algebra, the
associativity condition can be expressed in the following three
equivalent ways
\be
L_iL_j=f_{ijk}L_k,~~~~R_iR_j=f_{ijk}R_k,~~~~[R_i,L_j^T]=0
\ee
In particular we see that $L_i$ and $R_i$ are representations of
the algebra (regular representations).

\section{Integration Rules}

In the following, ${\cal A}$ will stay for an associative algebra
with identity. We will say that ${\cal A}$ is a {\em
self-conjugated} algebra, if the two regular representations
corresponding to the right and left multiplications are
equivalent, {\em i.e.}
\be
R_i=C^{-1}L_iC
\ee
In this case one has
\be
L_i\ket{Cx}=\ket{Cx} x_i,~~~~\ket{Cx}=C\ket x
\ee
We then define the integration over ${\cal A}$ by the requirement
\cite{casalbuoni}
\be
\int_{(x)}\ket{Cx}\bra x=1
\ee
The problem here is that by using the algebra product, we get
\be
\int_{(x)}C_{ij}x_jx_k=C_{ij}f_{jkl}\int_{(x)}x_l=\delta_{ik}
\label{problem}
\ee
but these are $(n+1)^2$ equations for the $n+1$ unknown
quantities $\int_{(x)}x_l$. However, by assuming $x_0=I$, and by
taking $k=0$, we get
\be
\int_{(x)}C_{ij}x_j=\delta_{i0}\longrightarrow
\int_{(x)}x_i=(C^{-1})_{i0}
\ee
It is not difficult to show that this solves all the eqs.
(\ref{problem}) \cite{integrale}.

\section{Examples}
\subsection{Paragrassmann Algebras}

A Paragrassmann algebra ${\cal G}_1^p$ of order $p$ is generated
by the symbol $\theta$, such that $\theta^{p+1}=0$, and its
elements are given by $x_i=\theta^i$, $i=0,\cdots, p$. The
product rule is simply $\theta^i\theta^j=\theta^{i+j}$. It is not
difficult to show \cite{casalbuoni} that the $C$ matrix exists
and that $C_{ij}=\delta_{i+j,p}$. This matrix has the properties
$C^T=C$ and $C^2=1$. Therefore we get
\be
\int_{(\theta)}\theta^i=\delta_{ip}
\ee
In particular, for $p=1$ (Grassmann algebra), we reproduce the
Berezin's integration rules.

\subsection{Matrix Algebras}

Consider the algebra ${\cal A}_N$ of the $N\times N$ matrices. A
basis is given by the set of matrices $\{e^{(nm)}\}$ such that
$e^{(nm)}_{ij}=\delta_i^n\delta_j^m$. The product rule is
$e^{(nm)}e^{(pq)}=\delta_{mp}e^{nq}$. Also in this case the $C$
matrix exists \cite{integrale} and it is such that under its
action $e^{(nm)}\to e^{(mn)}$, that is
$C_{(mn)(rs)}=\delta_{ms}\delta_{nr}$. It has the properties
$C^T=C$ and $C^2=1$. Since the identity is given by
$I=\sum_{n=1}^N e^{(nn)}$ the integration rule gives
\be
\int_{(e)}e^{(rs)}=\sum_{n}(C^{-1})_{(nn)(rs)}=\delta_{rs}
\ee
For an arbitrary matrix $A$, it follows
\be
\int_{(e)} A=\sum_{nm=1}^Na_{nm}\int_{(e)} e^{(nm)}=Tr(A)
\ee

\subsection{Projective Group Algebras}

Given a group $G$ and an arbitrary projective linear
representation ${\cal A}(G)$, $a\to x(a)$, with $a\in G$ and
$x(a)\in {\cal A}(G)$, the vector space ${\cal A}(G)$ has
naturally an algebra structure given by the product $x(a)x(b)=
\exp(i\alpha(a,b))x(ab)$, where the phase $\alpha(a,b)$ is the
cocycle associated to the projective representation. Also this
algebra is self-conjugated \cite{projective} with $C$ mapping
$x(a)$ into $x(a^{-1})$, that is $C_{ab}=\delta_{ab,e}$, where
$e$ the identity in the group. The integration rules are
\be
\int_{(x)}x(a)=\delta_{e,a}
\ee
In the case $G=R^D$, and ${\cal A}(G)$ a vector representation,
one has $x(\vec a)=\exp(i\vec q\cdot\vec a)$ and the algebraic
integration coincides with the integration over the variables
$\vec q$ \cite{projective}.

\section{Derivations}

There is an important theorem \cite{integrale} relating the
algebraic integration and a particular type of derivations on the
self-conjugated algebras which generalizes the ordinary
integration by part formula. The theorem says: {\em If $D$ is a
derivation such that $\int_{(x)}Df(x)=0$ for any function $f(x)$
on the algebra, then the integral is invariant under the related
automorphism $\exp(\alpha D)$, and viceversa}. If the matrix $C$
satisfies $C^T=C$, as in the examples above, it is also possible
to prove that the inner derivations give rise to automorphisms
leaving invariant the integration measure. The inner derivations
for an associative algebra are given by $D_a x_i=[x_i,a]$.
Therefore, for a matrix algebra our theorems implies
$\int_{(e)}D_A B=\int_{(e)}[B,A]=0$, which is nothing but
$Tr([B,A])=0$.

\section{Integration over a subalgebra}

Given a self-conjugated algebra ${\cal A}$ and a self-conjugated
subalgebra ${\cal B}$, we would like to recover the integration
over ${\cal B}$ in terms of the integration over ${\cal A}$.
Since we have the decomposition ${\cal A}={\cal B}\oplus{\cal
C}$, we look for an element $P\in {\cal A}$ such that $\int_{\cal
A}{\cal C}P=0$ and $\int_{\cal A}{\cal B}P=\int_{\cal B}{\cal
B}$, that is
\be
\int_{\cal A}{\cal A}P=\int_{\cal A}{\cal B}P=\int_{\cal B}{\cal B}
\ee
Since there are as many conditions as the elements of ${\cal A}$,
$P$ is uniquely determined. As an example we can consider a
Grassmann algebra ${\cal G}$ with a single generator $\theta$.
This can be embedded into the algebra ${\cal A}_2$, the algebra
of the $2\times 2$ matrices, by the correspondence
$\theta\to\sigma_+$, and $1\to 1_2$. Then, one finds $P=\sigma_-$
\cite{integrale}, and as a consequence one can express the
Berezin's integral in terms of traces over $2\times 2$ matrices
\be
\int_{{\cal A}_2}f(\sigma_+)\sigma_-=\int_{(\theta)}f(\theta)=
Tr[f(\sigma_+)\sigma_-]
\ee
Reproducing the known rules
\be
\int_{(\theta)}1=Tr[\sigma_-]=0,~~~~~\int_{(\theta)}\theta=
Tr[\sigma_+\sigma_-]=0
\ee

\section{Conclusions and Outlook}

We have shown that it is possible to define an integration
according to the principles outlined in the introduction for a
very large class of algebras (associative self-conjugated
algebras with identity). In particular, we have recovered many
known cases and shown that it is possible to generalize the
theorem of integration by part valid for the Lebesgue integral.
We have also noticed that the integration over subalgebras is
defined in a natural way. Furthermore, this approach can be
extended to algebras which are not self-conjugated (the right and
left multiplication representations are not equivalent), as the
algebra of the bosonic and of the $q$-oscillators
\cite{casalbuoni}. Also the case of the nonassociative octonionic
algebra can be treated by these methods \cite{casalbuoni}. The
next step in this approach would be the definition of the
path-integral via tensoring of the algebra.

\section*{Acknowledgments}

The author would like  to thank Prof. J. P. Eckmann, Director of
the Department of Theoretical Physics of the University of
Geneva, for the very kind hospitality.

\end{document}